\documentclass[12pt]{article}

\usepackage{graphicx}

\usepackage{graphicx}
\usepackage{booktabs}
\usepackage{ragged2e} 
\usepackage{supertabular}
\usepackage[dvipsnames]{xcolor}
\usepackage{hyperref}
\usepackage{caption}
\usepackage{subcaption}
\usepackage{float}

\definecolor{nblue}{RGB}{28,130,185}
\definecolor{cgreen}{RGB}{76,153,0}
\definecolor{myorange}{RGB}{245,156,74}

\usepackage{hyperref}
\hypersetup{
  colorlinks=true,
  citecolor=magenta,
  urlcolor=-myorange
}

\usepackage{amssymb}
\usepackage{amsmath}
\usepackage{amsfonts}
\newcommand{\be}{\begin{equation}}
\newcommand{\ee}{\end{equation}}
\newcommand{\bea}{\begin{eqnarray}}
\newcommand{\eea}{\end{eqnarray}}
\newcommand{\nn}{\nonumber}

\begin{document}

\title{Does a Curved Mirror Honestly Reflect Your Identity? A Study of Multipole Images in Front of a Grounded Sphere}

\author{Farhang Loran\thanks{E-mail address:
loran@iut.ac.ir} ~ and Saman Moghimi-Araghi\thanks{E-mail address: samanimi@sharif.edu }
\\[6pt]
$^{*}$Department of Physics, Isfahan University of Technology,\\  Isfahan 84156-83111, Iran\\
$^\dagger$Department of Physics, Sharif University of Technology, \\ Tehran, P.O. Box:11555-9161, Iran \\[6pt]
}

\date{ }
\maketitle

\begin{abstract}

The method of image charges is a powerful and elegant technique in electrostatics, commonly used to determine the electric field generated by point charges near conductors of various shapes. While standard problems focus on single charges interacting with conductors, the behavior of multipoles in such configurations has received comparatively less attention, particularly beyond the well-studied case of a flat plane. In this paper, we explore the image formation of electric dipoles and quadrupoles near a conducting sphere and uncover a wonderful result: the image of a given multipole is not necessarily of the same type. Instead, alongside the expected multipole image, the resulting image configuration also includes lower-order multipole contributions. This finding broadens the understanding of electrostatic images and offers new insights into the interaction of multipoles with conducting boundaries.

\vspace{6pt}

%We also comment on the application of our method for more general short-range potentials.

%\vspace{2mm}

%\noindent PACS numbers: 03.65.Nk, 42.25.Bs\vspace{2mm}


\end{abstract}
%\tableofcontents

\section{Introduction}

One of the fascinating exhibits sometimes displayed in science museums involves mirrors with multiple concave and convex surfaces. When people stand in front of these mirrors, they see distorted reflections of themselves—at times appearing fatter or thinner, with elongated or shortened legs. However, despite these distortions, they still recognize the reflection as their own image, albeit with some modifications. A similar concept can be explored in electrostatics through the method of images.

In the method of images, a charge or a set of charges is placed in front of a conducting surface, and the goal is to determine the resulting electric field outside the conductor \cite{Griffiths, Jackson}. To solve this problem, one should solve Poisson’s equation with the appropriate boundary conditions that enforce the equipotential nature of the conductor. While this can be approached using Green’s functions, an alternative method relies on the uniqueness theorem \cite{Jackson}: if a potential distribution satisfies Poisson’s equation and the given boundary conditions, it must be the unique solution. This allows us to construct a solution in certain symmetric geometries by introducing a set of fictitious ``image charges'' placed inside the conductor. These image charges are chosen so that the total field outside the conductor satisfies the required boundary conditions. Through this process, we determine the so-called image of a point charge in the presence of a conductor with a specific geometry.
Most textbooks focus on the images of single point charges or discrete sets of charges, with less attention given to more complex charge distributions such as multipoles. The special case of a dipole facing an infinite conducting plane is commonly discussed in introductory electrodynamics textbooks \cite{Griffiths}, but the analogous problem for a grounded conducting sphere is often overlooked. Nevertheless, some sources do address this configuration. For example, it appears in the textbook by Batygin and Toptygin \cite{Batygin}, and a particular case is solved in \cite{Grechko}.  A more general treatment is provided in \cite{Santos}, where the electric field and potential are calculated throughout space. However, these works primarily focus on obtaining the fields and potentials.   { In addition, in \cite{Bhatt}, the image of an arbitrary charge distribution near a conducting sphere has been formally derived. }
Similar configurations—such as the image of a magnetic dipole near a superconducting sphere—have also been studied \cite{Shah}. Moreover, image-charge techniques have been applied to compute the force between charged conducting spheres \cite{Sliško,Kolikov}. Nevertheless, the question of how a general multipole is imaged by a conducting sphere—specifically, what kind of multipole the image becomes—has received comparatively little attention.

The central question we address here is the following: what is the structure of the image when a multipole is placed in front of a conducting object such as a sphere? From an electrostatic perspective, different multipoles exhibit fundamentally distinct characteristics, particularly in their far-field behavior. The electric field of a pure $n$-pole falls off as $1/r^{n+1}$ at large distances, which may suggest that multipolarity is an intrinsic feature of a charge distribution. Given this, one might expect that placing a pure multipole in front of a conductor would yield another pure multipole as its image. While this expectation holds for a flat conducting plane, we show that for a grounded conducting sphere, the image of a multipole generally consists of a monopole and all lower-order multipoles up to the original one. 
  {In fact, when this question was asked of students who had taken electromagnetism courses and were familiar with the method of images, all initially answered—based on their experience with the flat conductor—that the multipolarity should be preserved. Only after a careful line of reasoning did they realize that this is not actually the case.} 

  {Moreover, most students have little intuition about multipole configurations beyond the dipole, even though quadrupole structures often arise in physical contexts. We found this problem to be a particularly suitable playground for pedagogically establishing a clear and graphical connection between various quadrupole charge configurations and their corresponding quadrupole tensors.}

The structure of this paper is as follows. In Section 2, we briefly review the method of image charges for a grounded conducting sphere. Section 3 is devoted to analyzing the image of a dipole, where we examine various configurations. In Section 4, we turn to the case of a quadrupole. Alongside a general analysis, we present specific examples that highlight situations where the image includes or excludes monopole and dipole components. Finally, in Section 5, we consider the case of an arbitrary multipole placed near the sphere, and demonstrate that its image generally includes lower-order multipoles.

\section{Image Charge Method for a Grounded Conducting Sphere}  

In this short section, we analyze the image charge problem for a single point charge in the presence of a grounded conducting sphere. This classic result will serve as the foundation for our later discussion of multipole images.  

\begin{figure}
    	\centerline{\includegraphics[width=0.48\textwidth]{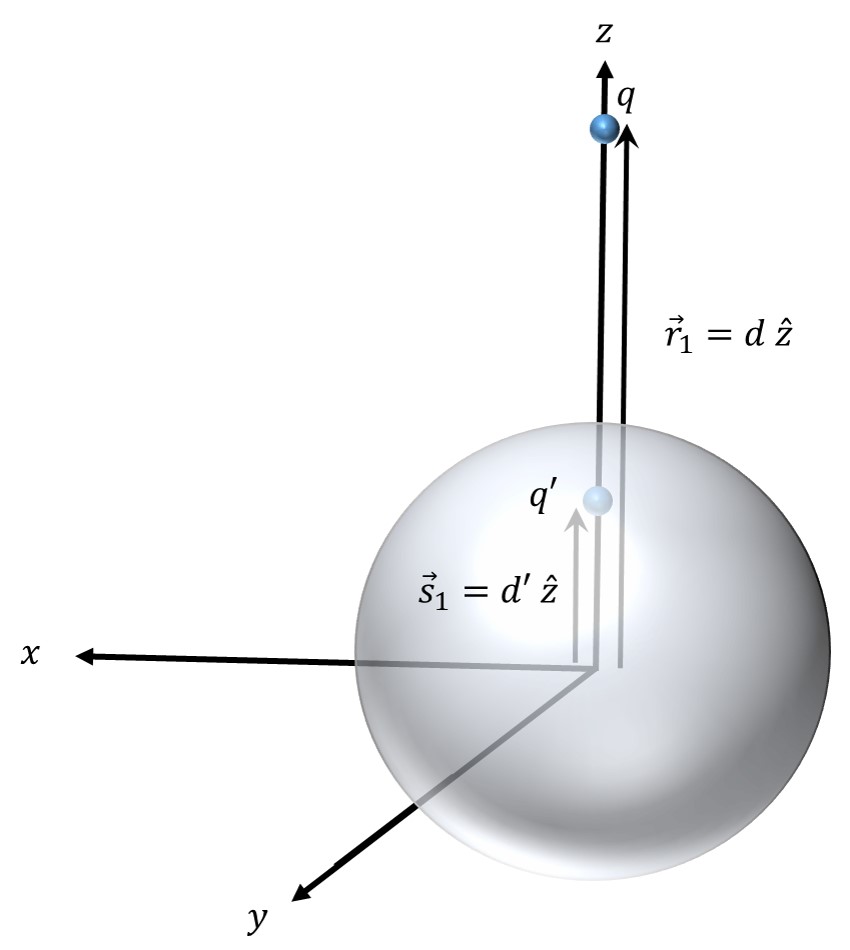}}
    
    \caption{The image of a single point charge in the vicinity of a grounded conduction sphere.}
    \label{fig:image}
\end{figure}

Consider a grounded conducting sphere of radius $R$, and a point charge $q$ placed at a distance $d$ from its center, with $d > R$ (see Figure (\ref{fig:image})). Due to the spherical symmetry of the problem, we can assume without loss of generality that the charge is placed along the $z$-axis at $\vec{r}_1=d\, \hat{z}$, where $\hat z$ denotes the unit vector pointing along the positive $z$ axis. To satisfy the boundary condition that the sphere remains an equipotential surface at zero potential, we replace the conductor with an image charge $q'$ located at $d'\, \hat{z}$ inside the sphere, where  

\begin{equation}
q' = -q \frac{R}{d},
\end{equation}
\begin{equation}
d' = \frac{R^2}{d}.
\end{equation}

The total potential outside the sphere is then given by the superposition of the potentials due to $q$ and $q'$:  

\begin{equation}
\label{monopole-potential}
\phi(\vec{r}) = \frac{1}{4\pi\epsilon_0} \left( \frac{q}{|\vec{r} - d \hat{z}|} + \frac{q'}{|\vec{r} - d' \hat{z}|} \right).
\end{equation}
This potential satisfies Poisson's equation at $r>R$ and is zero at the boundary $r=R$, where $r:=|\vec{r}|$. The corresponding electric field outside the sphere is obtained as  $\vec{E} = -\nabla \phi.$
This result will be useful later when analyzing the images of multipole sources in front of a grounded conducting sphere. The above solution,  {for $q=1$,} can be regarded as the Green's function for Poisson's equation with the specific boundary condition. The potential and electric field of more complicated charge distributions can be obtained simply by integration. 

Before turning to the problem of image dipoles and higher-order multipoles near a conducting sphere, it is helpful to briefly review the case of the image method in front of a very large conducting plane. For such a plane, the image charge is simply the mirror reflection of the original charge, with opposite sign. As a result, if we place a complex charge configuration near the plane, its image will reproduce the same structure on the other side of the plane, with all charges reversed in sign. This implies that the image of a monopole is a monopole, the image of a dipole is a dipole, and so on. Even if the source consists of a nontrivial combination of multipoles, the image preserves the same combination, mirrored across the plane.

In the next section, we will see that the situation is quite different for a conducting sphere. In that case, the image of a dipole may not remain a dipole—it can acquire components of different multipole order.

\section{Image of a dipole in front of a grounded sphere}

We aim to determine the image of an arbitrary electric dipole located near a conducting sphere. We set the origin of the coordinate system at the center of the sphere and align the $z$-axis such that the dipole is positioned on it. Various orientations of the dipole can be considered: it may be parallel to the $z$-axis, perpendicular to it, or form an angle \( \theta \) with it. However, due to the azimuthal symmetry of the problem, we can assume without loss of generality that the dipole lies in the \( xz \)-plane. To maintain a pedagogical and educational approach, we first consider the case where the dipole is perpendicular to the $z$-axis. Then, we examine the configuration where the dipole is parallel to the $z$-axis, and finally, we analyze the general case.

\subsection{Case 1: Dipole Perpendicular to the $z$-axis}  

Any dipole can be represented as a pair of point charges $ \pm q $ separated by a distance $ a $. To ensure a pure dipole and eliminate the effects of higher-order multipoles, we take the limits $ q \to \infty $ and $ a \to 0 $ while keeping $ qa = p $ constant.  

To obtain the image of the dipole in the presence of the conducting sphere, we first place two point charges $ \pm q $ at a separation $ a $, determine their respective image charges separately, and then take the aforementioned limits to analyze the resulting image structure. An initial expectation might be that the image remains a dipole.  

\begin{figure}
    	\centerline{\includegraphics[width=0.48\textwidth]{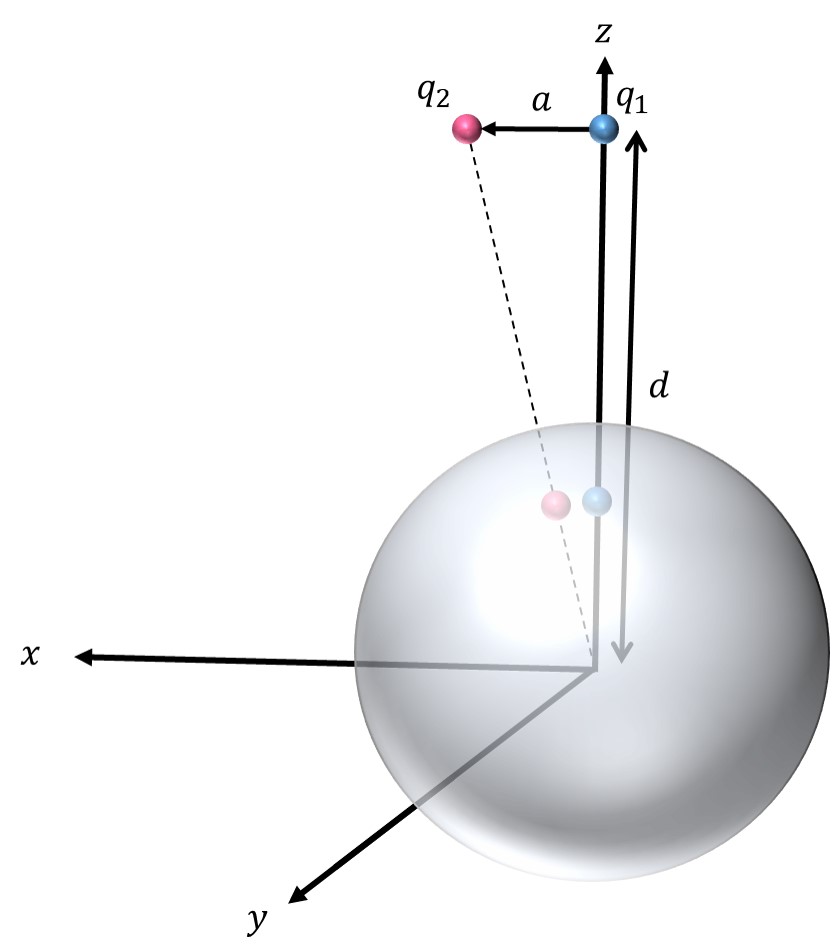}}
    
    \caption{The configuration of the dipole and the grounded conducting sphere in the first case.}
    \label{fig:dipole-1}
\end{figure}

For the first case, we arrange the charges as shown in Figure~(\ref{fig:dipole-1}). In this figure, the distance $a$ is exaggerated for clarity. Specifically, the position of charge $q_1 = -q$ is given by $\vec{r}_1 = d\hat{z}$, and the position of charge $q_2 = +q$ is given by $\vec{r}_2 = d\hat{z} + a\hat{x}$. This charge configuration forms a dipole of magnitude $\vec{p} = q a \hat{x}$ located at $\vec{r}_1$.  

To determine the image of the dipole, we first compute the image charges ($q_1'$ and $q_2'$) and their respective positions ($\vec{s}_1$ and $\vec{s}_2$) using equations (1) and (2):  

\begin{eqnarray}
q_1' &=& \frac{R q}{d},   \hspace{2cm}
q_2' = \frac{-R q}{\sqrt{d^2 + a^2}}.  \\
\vec{s}_1 &=& \frac{R^2}{d} \hat{z},  \hspace{2cm}
\vec{s}_2 = \frac{R^2}{d^2 + a^2} (d\hat{z} + a\hat{x}).
\end{eqnarray}
Since $\vec s_1$ is independent of $a$, we consider it as the location of the image dipole. Having the limits described earlier in mind, we only keep the leading terms and use the following approximations:
\begin{eqnarray}
q_2' \simeq -q_1',  \hspace{2cm}
\vec{s}_2 \simeq \frac{R^2}{d^2} (d\hat{z} + a\hat{x}).
\end{eqnarray}
Thus, the total image charge sums to zero, meaning no monopole contribution exists, and the only remaining structure is a dipole:
\begin{eqnarray}
\vec{p}\,' = q_1' \vec{s}_1 + q_2' \vec{s}_2  = -\frac{R^3}{d^3} q a \hat{x}
        = -\frac{R^3}{d^3} \vec{p}.
\end{eqnarray}

The fact that the image of a dipole remains a dipole is a satisfying and somewhat expected result, as we have seen the same behavior in the case of an infinite conducting plane. However, as we will see, this result does not hold in the general case, leading to more intriguing results.

\subsection{Case 2: Dipole Parallel to the $ z $-axis}

\begin{figure}
    	\centerline{\includegraphics[width=0.48\textwidth]{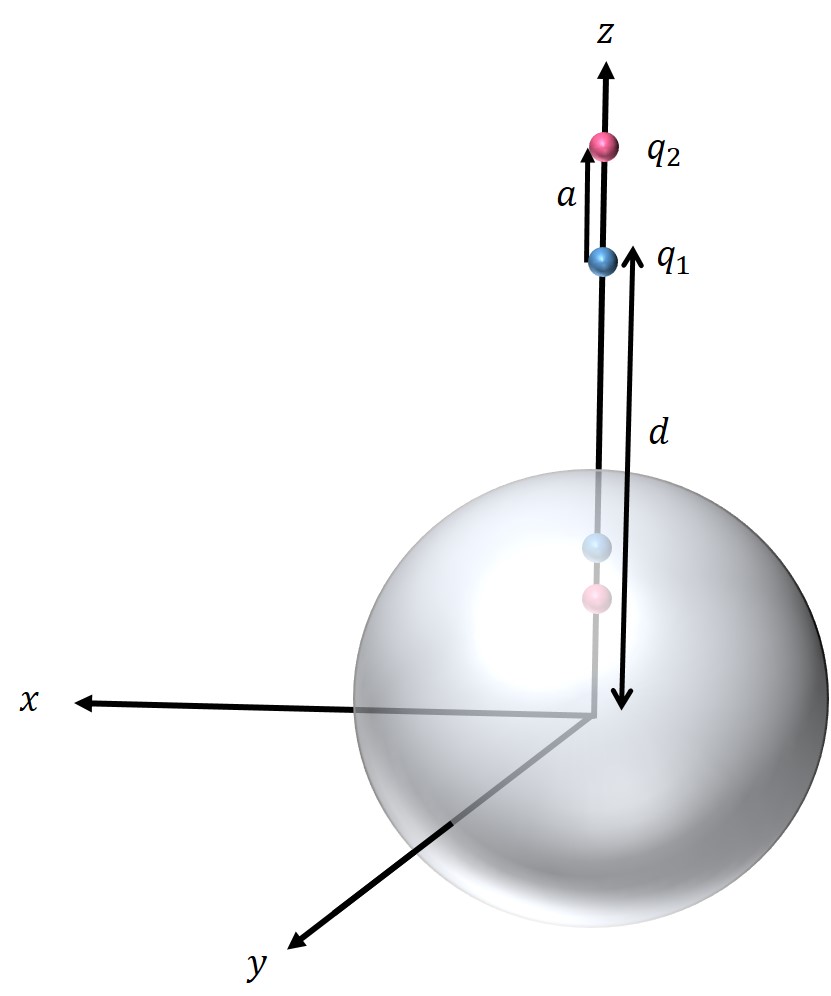}}
    
    \caption{The configuration of the dipole and the grounded conducting sphere in the second case.}
    \label{fig:dipole-2}
\end{figure}

Similar to the approach in the previous section, we construct a dipole by placing a charge $ q_1 = -q $ at $ \vec{r}_1 = d \hat{z} $ and another charge $ q_2 = +q $ at $ \vec{r}_2 = (d + a) \hat{z} $. Taking the limit where $ a \to 0 $ and $ q \to \infty $ while maintaining $ p:=q a $ finite, we obtain a dipole of magnitude $ \vec{p} = q a \hat{z} $, located at $ \vec{r}_1 = d \hat{z} $. The configuration of the charges and their corresponding image charges is illustrated in Figure \ref{fig:dipole-2}. The image resulting from this charge arrangement consists of two point charges located close to each other along the $z$-axis. This naturally suggests that the image itself might also form a dipole. Let us determine the image charges and examine whether our initial expectation is indeed correct.

The magnitudes of the image charges are given by
\begin{eqnarray}
\label{parallel-q1}
q_1' &=& \frac{R}{d} q, \\
\label{parallel-q2}
q_2' &=& -\frac{R}{d+a} q \simeq -\frac{R}{d} \left( 1 - \frac{a}{d} \right) q. 
\end{eqnarray}

Interestingly, these two charges are not equal up to first order in $a$. If we add them together, the total charge is proportional to $qa$ which is finite. Therefore, the image charge distribution contains a monopole component. This shows that in the presence of the conducting sphere, the nature of the image distribution has been fundamentally altered, now including a monopole component.

More precisely, the total image charge is
\begin{equation}
\label{parallel-Q}
Q = q_1' + q_2' = \frac{Rqa}{d^2} = \frac{R}{d^2} \, p,
\end{equation}
which decays as the inverse square of the distance, and for a given dipole moment, reaches an extremum of $p/R$.

Let us now compute the dipole moment of the resulting image. Since the total image charge is nonzero, the dipole moment depends on the choice of origin. However, to isolate the monopole and dipole components of the image distribution, we should evaluate the dipole moment with respect to the image location, namely $\vec{s}_1 = \frac{R^2}{d} \hat{z}$. Alternatively, one can compute the total dipole moment with respect to the origin and subtract the contribution from the monopole part accordingly.  {As shown in Appendix \ref{App-parallel-dipole}, the result is:}
\begin{equation}
\vec{p}\,' = \frac{R^3}{d^3} \, \vec{p}.
\end{equation}

Note that although this result is similar in magnitude to what we obtained in Case 1, the direction of the image dipole is now aligned with the original dipole, whereas in Case 1, they are antiparallel.

  { One may wonder why in the first configuration, no monopole appeared in the image, while in the second a monopole component emerged. This difference can be understood heuristically. The key point is that the magnitude of the image charge for a point charge depends on its distance from the center of the sphere. In the first configuration, the difference between the distances of the two charges from the center is of order $a^2$, whereas in the second configuration it is of order $a$. Consequently, in the limit $a \to 0$, the total image charge (i.e., the monopole component) vanishes in the first case but remains finite in the second. The same phenomenon will be observed in the next section when we consider the general case of dipole image. }

\subsection{The General Case: An Arbitrary Dipole Orientation}
%\section*{General Orientation of the Dipole}

In this section, we analyze the most general case, where the dipole is neither parallel nor perpendicular to the $z$-axis. Without loss of generality, we can assume that
\[
\vec{p} = p_1 \hat{x} + p_2 \hat{z}.
\]
Since the system is linear, the solution to this problem is simply the linear combination of the solutions obtained in the previous two cases. Specifically, the image charge distribution includes a monopole of magnitude
\[
Q = \frac{R}{d^2} p_2,
\]
located at
\[
\vec{s}_1 = \frac{R^2}{d} \hat{z},
\]
and a dipole given by
\[
\vec{p}\,' = \frac{R^3}{d^3} (-p_1 \hat{x} + p_2 \hat{z}).
\]

\begin{figure}
    	\centerline{\includegraphics[width=0.38\textwidth]{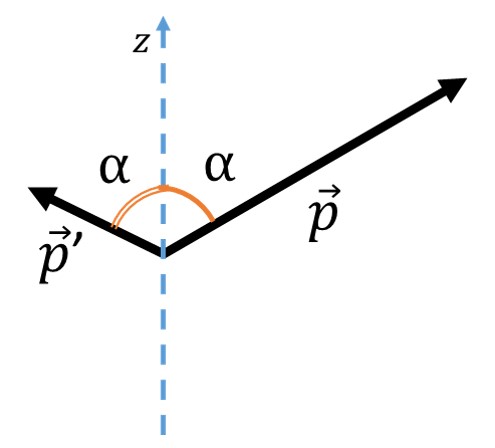}}
    
    \caption{  {An arbitrary dipole and its image dipole sketched together. Apart from the overall scaling factor, the $z$-component is preserved, while the $x$-component is mirrored with respect to the $z$–axis.} }
    \label{fig:dipoleandImage}
\end{figure}

  {In other words, up to an overall scaling factor, the image dipole has the same component along the $z$-axis but an opposite component along the $x$-axis, resulting in a reflection of its orientation with respect to the $z$–$x$ plane. (See Figure \ref{fig:dipoleandImage})}

As an interesting physical consequence, let us consider a dipole of fixed magnitude $p_0$ lying in the $xz$-plane, which is free to rotate around the $y$-axis. Let $\alpha$ be the angle between the dipole and the $z$-axis. The image of such a dipole consists of a monopole and a dipole located at the image point, both of which depend on the angle $\alpha$. The magnitude of the image dipole remains constant and is equal to $p_0 R^3/d^3$, but its direction flips, forming an angle $ - \alpha$ with the $z$-axis   {(Figure \ref{fig:dipoleandImage})}. In other words, as the original dipole rotates clockwise, the image dipole rotates counterclockwise. More intriguingly, the total image charge (monopole component) varies with $\alpha$, ranging within the interval
\[
Q \in \left[ -\frac{R}{d^2} p_0, \, \frac{R}{d^2} p_0 \right].
\]
  {This means that by simply rotating the dipole, the induced charge on the grounded sphere readjusts, as if electric charge were flowing into or out of the sphere through its grounding connection.} Recall that the sphere is held at zero potential, which effectively models it as being connected to infinity via a conducting wire. 

%\subsection*{Electric Potential Due to a Dipole in Front of a Conducting Sphere}

We can now compute the electric potential outside the sphere. This potential consists of three contributions: the original dipole located at $\vec{r}_1 = d \hat{z}$, and the monopole and dipole image charges, both located at $\vec{s}_1$. That is,
\begin{equation}
\phi(\vec{r}) = \phi_p(\vec{r}) + \phi_Q(\vec{r}) + \phi_{p'}(\vec{r}),
\end{equation}
where:
\begin{align}
\phi_p(\vec{r}) &= \frac{1}{4\pi\epsilon_0} \frac{\vec{p} \cdot (\vec{r} - \vec{r}_1)}{|\vec{r} - \vec{r}_1|^3}, \\
\label{dipole-potential-Q}
\phi_Q(\vec{r}) &= \frac{1}{4\pi\epsilon_0} \frac{Q}{|\vec{r} - \vec{s}_1|}, \\
\label{dipole-potential-p}
\phi_{p'}(\vec{r}) &= \frac{1}{4\pi\epsilon_0} \frac{\vec{p}\,' \cdot (\vec{r} - \vec{s}_1)}{|\vec{r} - \vec{s}_1|^3}.
\end{align}

 {Appendix~\ref{App-dipole} illustrates how these terms can be combined and simplified to produce}
\begin{equation}
\label{dipole-potential}
\phi(\vec{r}) = \frac{1}{4\pi\epsilon_0} \, \vec{p} \cdot \left( \frac{\vec{r} - \vec{r}_1}{|\vec{r} - \vec{r}_1|^3} + \frac{\frac{r^2}{R^2} \vec{r}_1 - \vec{r}}{\left| \frac{d}{R} \vec{r} - R \hat{z} \right|^3} \right).
\end{equation}

There are a few noteworthy features regarding the potential obtained. First, observe that the potential decays as \( r^{-1} \) at large distances, which reflects the presence of a monopole component in the image distribution. Second, the potential should satisfy the boundary condition of vanishing exactly on the surface of the sphere. To verify this, we consider an arbitrary point \( \vec{R} \) on the sphere. Naturally, the magnitude of this vector is \( R \). Noting that \( |\vec{R} - d \hat{z}| = |d \hat{R} - R \hat{z}| \),  {with $\hat{R}:=R^{-1}\vec R$,} one readily finds-- {see Appendix \ref{App-dipole}}--that \( \phi(\vec{R}) = 0 \) for all points on the sphere. Hence, the potential satisfies the required boundary condition.

Another important point is that the same potential can be derived using an entirely different method: the Green’s function approach. As previously noted, the method of images effectively constructs the Green’s function for the Laplace operator subject to specific boundary conditions; in this case, zero potential on the grounded conducting sphere. Given any charge distribution, the corresponding potential can then be computed by integrating the Green’s function against the source.  {In Appendix \ref{Appendix-charge-distribution}, we show that the charge distribution corresponding to a dipole is (see \cite{Jackson} Problem 4.2):}
\begin{equation}
\label{dipole-density}
\rho(\vec{r}\,') = -\vec{p} \cdot \nabla' \delta(\vec{r}\,' - \vec{r}_1),
\end{equation}
where $\nabla'$ is the derivative with respect to $\vec r\,'$ and $\vec{r}_1$ is the location of the dipole. Eq.~\eqref{monopole-potential} implies that the Green’s function for a unit point charge in the presence of a grounded conducting sphere is:
\begin{equation}
\label{Green}
G(\vec{r}, \vec{r}\,') = G_{\text{ch}}(\vec{r}, \vec{r}\,') +G_{\text{im}}(\vec{r},\vec{r}\,') 
\end{equation}
 where 
 \begin{equation}
G_{\text{ch}}(\vec{r}, \vec{r}\,') := \frac{1}{4\pi\epsilon_0}  \frac{1}{|\vec{r} - \vec{r}\,'\,|},
\end{equation}
and 
\begin{equation}
\label{Green-image}
    G_{\text{im}}(\vec{r}, \vec{r}\,'):= - \frac{1}{4\pi\epsilon_0}\frac{ R/ r'}{\left| \vec{r} - \left( \frac{R}{r'} \right)^2 \vec{r}\,' \,\right|},
\end{equation}
in which, $r'$ denotes the magnitude of the vector $\vec{r}\,'$.
To obtain the potential due to the dipole, we evaluate:
\begin{equation}
\phi(\vec{r}) = \int d^3r' \, \rho(\vec{r}\,') G(\vec{r}, \vec{r}\,').
\end{equation}
Integration by parts moves the derivative from the delta function to the Green’s function, giving:
\begin{equation}
\label{dipole-potential-Green}
\phi(\vec{r}) = \vec{p} \cdot \nabla' \left.G(\vec{r}, \vec{r}\,')\right|_{\vec{r}\,'=\vec{r}_1},
\end{equation}
which immediately recovers the expression obtained earlier in \eqref{dipole-potential}.  {Specifically, as demonstrated in Appendix \ref{App-Green-dipole},} the second term on the right-hand side of that equation, which represents the electric potential due to the image charge and image dipole, is given by
\begin{equation}
\phi_{\text{im}}(\vec{r}) = \vec{p} \cdot \nabla'\left. G_{\text{im}}(\vec{r}, \vec{r}\,')\right|_{\vec{r}\,'=\vec{r}_1}.
\end{equation}

As a final topic in our discussion of a dipole near a conducting sphere, let us examine its electrostatic energy. To calculate this quantity, we need to evaluate the electric field of the electric field of the image charge and the image dipole $\vec{E}_{\text{im}}(\vec{r}):=-\nabla\phi_{\text{im}}(\vec r)$, at $\vec r=\vec r_1$  where the original dipole is located. 
\begin{equation}
    \vec{E}_{\text{im}}(\vec{r}_1)= \frac{R}{4\pi\epsilon_0} \, \frac{(\vec{p} \cdot \vec{r}_1)\vec{r}_1 + R^2 \vec{p}}{(r_1^2 - R^2)^3}.
\end{equation}
The energy is simply given by
\begin{eqnarray}
    U &=& -\vec{p} \cdot \vec{E}_{\text{im}}(\vec{r}_1)\nn\\
&=& -\frac{R}{4\pi\epsilon_0} \,\frac{(\vec{p} \cdot \vec{r}_1)^2 + R^2 p^2}{(r_1^2 - R^2)^3}.
\end{eqnarray}

Note that since the sphere is held at zero potential—equivalent to the potential at infinity—adding or removing net charge to the sphere does not require any work. Another important point is that this energy decays as $r_1^{-4} = d^{-4}$ at large distances. This scaling originates from the fact that at large distances, we can neglect the part of the field which is due to the dipole in the image and just keep the monopole part. In addition, the magnitude of the image monopole decreases as $d^{-2}$, and the field generated by this monopole also decays as $d^{-2}$. The product of these two effects results in an electric field at the dipole’s location that scales as $d^{-4}$, which is exactly what appears in the above expression for the energy.

\section{Image of Quadrupoles in a Grounded Conducting Sphere}\label{Section-4.1}

We have already seen how the image of a dipole placed near a grounded conducting sphere may include a monopole component—a surprising outcome given our familiarity with image methods in simpler geometries like a grounded plane. This naturally leads to a deeper question: what happens if we place a quadrupole in front of such a sphere? Will the image remain purely quadrupolar? Or might it include dipolar or even monopolar terms?

To investigate this, we take a more general approach. We know the image of a single point charge located at $\vec{r}$ appears at the inverted point $\vec{r}\,' := \dfrac{R^2}{r^2} \vec{r}$ with rescaled magnitude. If we are given a continuous charge distribution $\rho(\vec{r})$, we can ask what the image charge distribution $\rho'(\vec{r}\,')$ will look like. This requires not only accounting for the transformation of the charge magnitude but also including the Jacobian of the spatial inversion.

To make this approach concrete, we recall that the method of images gives an explicit rule for computing the image of a general continuous charge distribution. Suppose \( \rho(\vec{r}) \) is localized outside the conducting sphere around the point $\vec r_1$. Define the infinitesimal source charge element as
\begin{equation}
\label{dq}
dq(\vec{r}) := \rho(\vec{r})\, d^3r.
\end{equation}
The corresponding image charge element is a charge element inside the sphere given by:
\begin{equation}
\label{dq'}
dq'(\vec{r}\,') = - \frac{R}{|\vec{r}\,|} \, dq(\vec{r}), \qquad 
\vec{r}\,' = \frac{R^2}{|\vec{r}\,|^2} \vec{r}.
\end{equation}
This is the so-called Kelvin transformation \cite{Kelvin}, which maps the region outside the sphere to the interior via inversion, and rescales the charge density appropriately to maintain the correct potential. This transformation has applications in multipole expansions in electrostatics \cite{Amaral}.

This transformation can be used to compute the image charge distribution explicitly. But to compute the corresponding multipoles, it suffices to note that this charge distribution is localized around 
    \begin{equation}
    \label{s1}
    \vec{s}_1:= \frac{R^2}{|\vec{r_1}|^2} \vec{r}_1.    
    \end{equation}
Thus, the corresponding monopole, dipole, and quadrupole are given by
\begin{align}
\label{image-charge-general}
q' &= \int dq', \\
\label{image-dipole-general}
p'_i &= \int dq'\, \Delta r'_i \\
\label{image-quad-general}
Q'_{ij} &= \int dq'\, \left(3\Delta r'_i \Delta r'_j - |{\Delta \vec r}\,'|^2 \delta_{ij}\right),
\end{align}
where $\Delta \vec r\,':=\vec r\,'-\vec s_1$,  {and $\delta_{ij}$ is the Kronecker delta}. Higher-order multipoles can be calculated similarly. 

Although this procedure can be carried out for an arbitrary charge distribution, we focus on the specific case of the image of a quadrupole. The charge distribution of an ideal quadrupole located at $\vec{r}_1$ is given by
\begin{equation}
\label{quadrupole-density}
\rho(\vec{r}) = \frac{1}{6} Q_{ij} \, \partial_i \partial_j \delta(\vec{r} - \vec{r}_1),
\end{equation}
where $Q_{ij}$ is a symmetric, traceless tensor representing the quadrupole moment,  {\[\partial_i:=\frac{\partial}{\partial x_i},\] and $x_i:=\vec r\cdot\hat e_i$ denotes the $i$-th component of the vector $\vec r$.}\footnote{ {This observation can be substantiated via a method analogous to that used in the dipole case, as presented in Appendix \ref{Appendix-charge-distribution}}.}

 {As demonstrated in Appendix~\ref{App-images}, substituting \eqref{quadrupole-density} into \eqref{dq} and applying \eqref{dq'} yields, via equations \eqref{image-charge-general}, \eqref{image-dipole-general}, and \eqref{image-quad-general}, the following expressions:}
\begin{eqnarray}
\label{monopol4im}
q' &=& -\frac{R}{2} \, Q_{ij} \, \frac{(r_1)_i (r_1)_j}{|\vec{r}_1|^5},\\
\label{dipole4im}
\vec{p}\,' &=& R^3 Q_{ij} \frac{(r_1)_i}{|\vec{r}_1|^5} \hat{e}_j
+ \frac{4R^2}{|\vec{r}_1|^2} \, q' \vec{r}_1,\\
\label{quadrapole4im}
Q'_{ij} &=& -\frac{R^5}{|\vec{r}_1|^5} Q_{ij}
+ \frac{2R^5}{|\vec{r}_1|^7} \Big( Q_{ik} (r_1)_k (r_1)_j + i \leftrightarrow j \Big)
- \frac{4R^5}{|\vec{r}_1|^9} \Big( Q_{kl} (r_1)_k (r_1)_l \Big) (r_1)_i (r_1)_j,
\end{eqnarray}
 {where $\hat e_i$ represents the unit vector pointing along the $i$-th axis.}
It is clear that $q'$ and $\vec p\,'$ are non-zero in general, confirming that even a pure quadrupole placed outside the sphere can induce monopolar and dipolar terms in its image.

This is a sharp departure from the behavior observed in flat geometries and highlights the rich structure of image multipoles in spherical systems. In the following subsection, we explore a few illustrative examples of quadrupole configurations and analyze how the monopole and dipole components in their images vary with geometry and orientation.

These expressions exhibit a clear hierarchical structure:   {noting $\vec{r}_1=d\,\hat{z}$, with $d$ being the distance of the original quadrapole to the centre of the sphere, in is observed that} the image of a quadrupole induces a monopole of order \((R/d)^3\), a dipole of order \((R/d)^4\), and a quadrupole of order \((R/d)^5\), all with coefficients involving appropriate contractions with \(Q_{ij}\). This scaling is not only consistent with the multipolar structure but also required by dimensional analysis: to generate a dipole from a quadrupole distribution, one needs a length scale to contract with \(Q_{ij}\), resulting in terms of the form \( Q_{ij}/R\); similarly, a monopole contribution must arise from a contraction involving two length scales, producing terms like \( Q_{ij}/R^2\).   {In addition, the origin of the dependence on powers of $(R/d)$ can be understood from another point of view. As shown in~\cite{Amaral}, if a charge distribution with density $\rho$ is placed near a conducting sphere, the image charge distribution is given by $\rho' = -( \frac{R}{d})^5 \rho$.
This relation is essentially responsible for the appearance of these coefficients in the above expansion. Of course, obtaining the complete result requires carrying out the detailed calculations presented earlier.} Moreover, all terms transform correctly under spatial rotations, and no other structures are possible that respect the   {covariance under rotations} and dimensional constraints of the problem.

\subsection{Specific Examples}

Although in the previous section we derived the general expressions for the image multipoles of an arbitrary quadrupole placed near a grounded conducting sphere, specific examples can offer more physical intuition and help clarify how different components of the quadrupole contribute to the image. For instance, one may ask under what conditions the image monopole or dipole vanishes.

To simplify the analysis, we assume that the quadrupole is located at $\vec{r}_1 = d \hat{z}$. With this assumption, the general expressions for the image monopole, dipole, and quadrupole, given in Eqs.~\eqref{monopol4im}, \eqref{dipole4im}, and \eqref{quadrapole4im} respectively, simplify to:

\begin{equation}\label{monoquad}
q' = -\frac{R}{2} \frac{Q_{zz}}{d^3},
\end{equation}

\begin{equation}\label{dipolequad}
\vec{p}\,' = \sum_{j=1}^3 \frac{R^3}{d^4} Q_{zj} \hat{e}_j + \frac{4R^2 q'}{d} \hat{z},
\end{equation}

\begin{equation}
Q'_{ij} = -\frac{R^5}{d^5} \left( Q_{ij} - 2 Q_{zi} \delta_{zj} - 2 Q_{zj} \delta_{zi} + 4 Q_{zz} \delta_{iz} \delta_{jz} \right).
\end{equation}

As these expressions make clear, the components $Q_{iz}$, and especially $Q_{zz}$, play a central role. In particular, the image monopole exists if and only if $Q_{zz} \neq 0$. Therefore, in order for both the image monopole and dipole to vanish, the third row and third column of the original quadrupole tensor must be zero.

We now construct quadrupoles with this property. In general, a quadrupole can be built from two opposing dipoles separated by a small distance. More concretely, we place four point charges—two positive and two negative—at a distance of order $a$, with $a \to 0$ and $q \to \infty$, keeping $qa^2 = Q$ constant. All of the following examples are constructed using this method.

\begin{figure}
    \centering
    \includegraphics[width=0.8\linewidth]{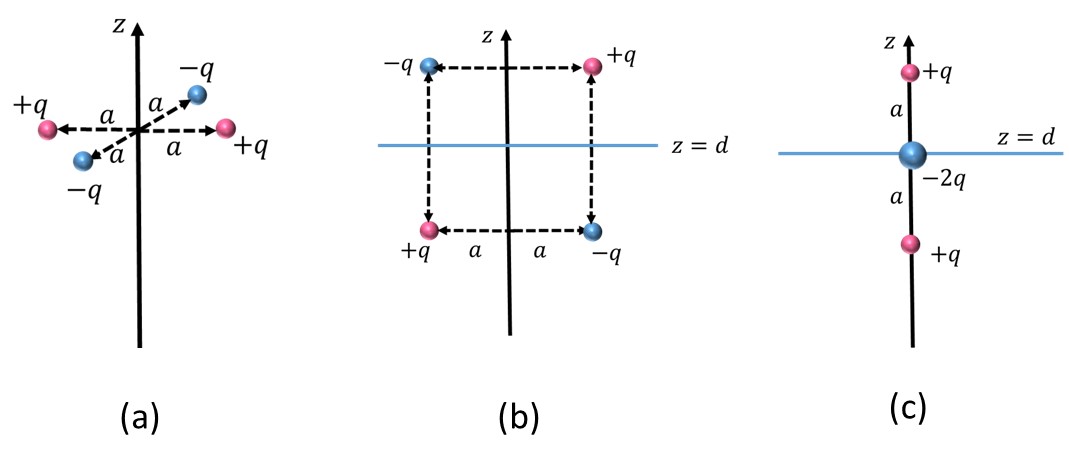}
    \caption{Charge configurations corresponding to Examples 1, 2, and 3, illustrated in panels (a), (b), and (c).}
    \label{fig:Quadpos}
\end{figure}

\paragraph{Example 1.}
All four charges lie in the plane $z = d$. Two positive charges are placed at $( \pm a, 0, d )$ and two negative charges at $( 0, \pm a, d )$ (see Figure \ref{fig:Quadpos}a). This quadrupole can be constructed in the following way: we begin with a dipole oriented perpendicular to the $z$-axis, and place alongside it an opposite dipole—equal in magnitude but reversed in direction—displaced by a small distance $\sqrt{2}a$ within the same $xy$-plane. In other words, not only are both initial dipoles perpendicular to the $z$-axis, but the relative displacement between them also lies entirely within the transverse plane.  {As illustrated in Appendix \ref{App-Quad-Example1},} the resulting quadrupole tensor is

\begin{equation}
\label{example 1}
Q_{ij} = 6qa^2  
\begin{pmatrix}
1 & 0 & 0 \\
0 & -1 & 0 \\
0 & 0 & 0
\end{pmatrix},
\end{equation}
which is traceless and whose third row and third column vanish. As a result, both the image monopole and dipole vanish, and the image consists purely of a quadrupole 
    \begin{equation}
        Q'_{ij}=-\frac{R^5}{d^5} Q_{ij}.
    \end{equation}

This situation mirrors the first dipole example discussed earlier, in which the dipole was perpendicular to the $z$-axis and the image consisted of a dipole only. Similarly here, the quadrupole lies entirely in the transverse plane and produces a purely quadrupolar image.

\paragraph{Example 2.}
Now, consider a quadrupole built from four point charges, placed as follows:
\[
\begin{aligned}
\vec{r}_I &= (a,0,d+a), & q_I &= +q,\\
\vec{r}_{II} &= (-a,0,d-a), & q_{II} &= +q,\\
\vec{r}_{III} &= (-a,0,d+a), & q_{III} &= -q,\\
\vec{r}_{IV} &= (a,0,d-a), & q_{IV} &= -q.
\end{aligned}
\]
This configuration can be interpreted as follows: we begin with a dipole oriented within the $xy$-plane, and then add an opposite dipole, equal in magnitude but reversed in direction, displaced along the $z$-axis by a small distance $2a$ (see Figure \ref{fig:Quadpos}b). In other words, while the dipoles themselves lie in the transverse plane, the relative displacement between them is purely vertical.
Taking the limit \( a \to 0 \), \( q \to \infty \) with \( q a^2 = Q \) held fixed, one finds that the only nonzero component of the quadrupole tensor is
\begin{equation}
Q_{xz} = Q_{zx} = 12\,q\,a^2 = 12\,Q,
\end{equation}
so that
\begin{equation}
Q_{ij} = 12\,Q\;
\begin{pmatrix}
0 & 0 & 1 \\
0 & 0 & 0 \\
1 & 0 & 0
\end{pmatrix}.
\end{equation}
Since \( Q_{zz} = 0 \), the image monopole
\begin{equation}
q' = -\frac{R}{2}\,\frac{Q_{zz}}{d^3} = 0,
\end{equation}
and because \( Q_{zy} = 0 \), only \( Q_{zx} \) survives in Eq. \eqref{dipolequad} and we have:
\begin{equation}
\vec{p}\,' 
= \frac{12QR^3}{d^4}\, \hat{x}.
\end{equation}
Thus, the image distribution exhibits a nonzero dipole moment along \( \hat{x} \), while the monopole component vanishes. As expected, the image also includes a quadrupolar term, which can be computed from Eq.~(\ref{quadrapole4im}). 
 \begin{equation}
        Q'_{ij}=\frac{R^5}{d^5} Q_{ij}.
    \end{equation}
\paragraph{Example 3.}
As a final example, consider a symmetric configuration in which all charges lie on the $z$-axis:
\[
\begin{aligned}
\vec{r}_I &= (0,0,d+a), & q_{I} &= +q, \\
\vec{r}_{II} &= (0,0,d-a), & q_{II} &= +q, \\
\vec{r}_{III} &= (0,0,d),   & q_{III} &= -2q.
\end{aligned}
\]

This quadrupole can be viewed as two vertical dipoles of equal and opposite strength, aligned along the $z$-axis and separated by a small distance $a$ (Figure \ref{fig:Quadpos}c). Thus, both the dipoles and their relative displacement lie entirely along the $z$ direction.

Taking the limit $a \to 0$, $q \to \infty$ with $qa^2 = Q/2$ fixed, the resulting quadrupole tensor is diagonal and traceless:
\begin{equation}
Q_{ij} =  Q 
\begin{pmatrix}
-1 & 0 & 0 \\
0 & -1 & 0 \\
0 & 0 & 2
\end{pmatrix}.
\end{equation}
Since $Q_{zz} = 2Q$ is nonzero, the image distribution includes a monopole:
\begin{equation}
q' =  -\frac{R Q}{d^3}.
\end{equation}
The dipole component of the image is also aligned along the $z$-axis and given by:
\begin{equation}
\vec{p}\,' = -\frac{2 Q R^3}{d^4} \hat{z}.
\end{equation}
and the image quadrapole is 
     \begin{equation}
        Q'_{ij}=-\frac{R^5}{d^5} Q_{ij}.
    \end{equation}
As expected from the full axial symmetry of the original configuration, all image multipole components are aligned with the $z$-axis. This is the only example among the third in which all three image moments -- monopole, dipole, and quadrupole -- are present.

One might wonder whether it is possible to construct a quadrupole whose image includes a monopole and a quadrupole component, but no dipole. Since we want the image to contain a monopole, equation (\ref{monoquad}) tells us that we must have $ Q_{zz} \neq 0 $. Let us now examine equation (\ref{dipolequad}), which gives the $z$-component of the image dipole. Substituting in the expressions, we find:
\begin{equation}
p'_z = \frac{R^3}{d^4} Q_{zz} + \frac{4R^2 q'}{d} = \frac{R^3}{2d^4} Q_{zz},
\end{equation}
where we have used the explicit expression for $q'$ from equation (\ref{monoquad}). As long as $Q_{zz} \neq 0$, the dipole moment cannot vanish. Therefore, it is not possible to design a quadrupole whose image contains only monopole and quadrupole terms while completely suppressing the dipole.

\subsection{Potential Due to a Quadrupole}

In this short section, we derive the electrostatic energy of a fixed quadrupole placed near a grounded conducting sphere. 
Using the Green's function \eqref{Green} and the charge distribution \eqref{quadrupole-density}, we verify that the potential has two terms. The potential of the quadrupole and the potential sourced by the image charge, image dipole, and image quadrupole together.  {As derived in Appendix~\ref{App-Green-image-quadrupole}, the latter takes the form}
    \begin{eqnarray}
\phi_{\text{im}}(\vec{r}) &:= &\int d^3r' \, \rho(\vec{r}\,)\, G_{\text{im}}(\vec{r}, \vec{r}\,')\nn\\ 
\label{image-quadrupole-potential}&=& -\frac{R}{8\pi\epsilon_0}Q_{ij}\frac{\Big(r^2(r_1)_i-R^2x_i\Big)\Big(r^2(r_1)_j-R^2x_i\Big)}{\Big(d^2r^2+R^4-2R^2\vec r\cdot\vec r_1\Big)^{5/2}}.
    \end{eqnarray}
The energy of a quadrupole \( Q_{kl} \) in the external field \( \vec{E}_{\text{im}}:=-\nabla\phi_{\text{im}} \) is given by

\begin{equation}
U = -\frac{1}{6} Q_{kl} \, \partial_k (E_{\text{im}})_l.
\end{equation}
While the computation of this quantity is straightforward, it is algebraically involved. Instead of carrying out the full derivation here, we focus on the structure of the resulting terms and their asymptotic behavior.

The field at the quadrupole's position is produced by the image monopole, image dipole, and image quadrupole. The image monopole is proportional to \( R/d^3 \), and its field behaves as \( 1/d^2 \) at the source point. Its gradient then scales as \( 1/d^3 \), leading to an energy contribution proportional to \( R/d^6 \). Similarly, the dipole and quadrupole contributions to the energy scale as \( R^3/d^8 \) and \( R^5/d^{10} \), respectively. 

The full expression for the electrostatic energy is the following:

\begin{eqnarray}
U = -\frac{R}{24\pi\epsilon_0 \left( r_1^2 - R^2 \right)^5}
\left[
\frac{3}{2} \left( Q_{ij} (r_1)_i (r_1)_j \right)^2
+6 R^2 (r_1)_i (r_1)_j Q_{ik} Q_{kj}
+ R^4 Q_{kl} Q_{kl}
\right],
\end{eqnarray}
where \( \vec{r}_1 =d\, \hat{z}\) denotes the position of the original quadrupole. This expression makes explicit the contributions from all multipole interactions and exhibits the asymptotic behavior discussed above.
At large distances \( r_1=d \gg R \), the term involving \( \left( Q_{ij} (r_1)_i (r_1)_j\right)^2 \) dominates, reflecting the leading effect of the image monopole on the quadrupole's energy.

\section{Higher-Order Multipoles}

Our ability to solve the quadrupole case encourages us to explore the more general problem of arbitrary higher-order multipoles. The natural question is: if a higher-order multipole is placed near a grounded conducting sphere, what kind of image distribution does it produce?

Based on the examples analyzed thus far, we can expect that, in general, the image of a multipole will incorporate contributions from all lower-order multipoles. 
To extend the approach used in the previous sections, one might attempt to place a general multipole near the conducting sphere and directly compute its image to see which multipoles are induced. However, this method becomes impractical as the order of the source multipole increases.
Instead, we adopt a more systematic and efficient strategy. Suppose we are given an arbitrary localized charge distribution \( \rho(\vec{r}) \) centered near the point \( \vec{r}_1 \). This distribution can be expanded in terms of its multipole moments around \( \vec{r}_1 \), and we may compute each moment explicitly. We then ask a different question: what is the monopole moment of the image distribution? More specifically, how does each original multipole moment contribute to the resulting monopole? The same question can be posed for the image dipole, quadrupole, and higher-order components.

This approach has the advantage of focusing on the image multipoles one at a time, starting with the monopole. If we can demonstrate that all source multipoles generically contribute to the image monopole, then we may conclude that a multipole of order \( n \) can, in general, induce image multipoles of all orders less than or equal to \( n \).

We thus begin with a charge distribution \( \rho(\vec{r}) \) localized near the point \( \vec{r}_1 \). Its monopole moment is given by:
\begin{equation}
\label{monopole-moment}
    q = \int d^{3}r\, \rho(\vec{r}),
\end{equation}
and its higher-order multipole moments are defined by:
\begin{equation}
\label{multipole-moment}
    Q_{i_1 \cdots i_n} := \int d^3r \, \rho(\vec{r}) \, T_{i_1 \cdots i_n}(\vec{r} - \vec{r}_1),
\end{equation}
where \( T_{i_1 \cdots i_n}(\vec{y}) \) denotes the symmetric, traceless Cartesian multipole tensor of rank \( n \), constructed from the components of the vector \( \vec{y} = \vec{r} - \vec{r}_1 \). For example, the first three tensors are:

\begin{align}
\label{Ti}
T_i(\vec{y}) &= y_i, \\
\label{Tij}
T_{ij}(\vec{y}) &= 3 y_i y_j - |\vec{y}|^2 \delta_{ij}, \\
\label{Tijk}
T_{ijk}(\vec{y}) &= 15y_i y_j y_k - 3 |\vec{y}|^2 ( y_i \delta_{jk} + y_j \delta_{ik} + y_k \delta_{ij} ),
\end{align}
 {where $y_i$ denotes the $i$-th component of the vector $\vec y$.}

Having the charge distribution at hand, we can compute its image, and then systematically decompose that image into standard multipole components located at $\vec s_1$ defined in \eqref{s1}. 
For example, let us compute the image monopole corresponding to a charge distribution described by a charge density $\rho(\vec r)$. The image monopole is given by \eqref{image-charge-general}.
\begin{equation}
q' = -R \int d^3 r\, \frac{\rho(\vec{r})}{|\vec{r}|}.
\end{equation}

Substituting the multipole expansion into this expression and integrating term by term, we obtain   {(see~\cite{Jackson} Section 4.1)}:
\begin{equation}\label{qgen}
q' = -\frac{R}{|\vec{r}_1|} \left( q 
+ \sum_{n=1}^\infty \frac{(-1)^n}{n!} \frac{(r_1)_{i_1} \cdots (r_1)_{i_n}}{|\vec{r}_1|^{2n}} Q_{i_1 \cdots i_n} \right).
\end{equation}

This equation indicates that, in general, all source multipoles contribute to the monopole term in the image. In other words, a multipole of arbitrary order induces a monopole component in its image. Likewise, contributions to higher-order image multipoles can be determined using \eqref{image-dipole-general}, \eqref{image-quad-general}, or their corresponding extensions for higher-order multipoles. 

One can also verify that the image charge of a multipole of order \( n \) does not contain multipoles of order higher than \( n \). We already know that the image of a point charge is simply a point charge. This can be confirmed by noting that the image multipole charge is given by 
    \begin{equation}
        Q'_{i_1 \cdots i_n} = \int dq' T_{i_1 \cdots i_n}(\vec{r}\,' - \vec{s}_1).
    \end{equation}
Using $\rho(\vec r)=q\delta(\vec r-\vec r_1)$ together with \eqref{dq} and \eqref{dq'} we verify that 
    \begin{eqnarray}
        Q'_{i_1 \cdots i_n} &=& -\frac{qR}{r}\left.T_{i_1 \cdots i_n}(\vec u)\right|_{\vec u=\vec 0}\nn\\
        &=&0,
    \end{eqnarray}
where $\vec u:=\frac{R^2}{r^2}\vec{r} - \frac{R^2}{|\vec r_1|^2}\vec{r_1}$.\footnote{ {This can be readily confirmed for the lowest-rank multipoles by directly using \eqref{Ti}--\eqref{Tijk}.}}
Similarly, for a dipole we can use $\rho(\vec r)=-\vec p\cdot\nabla\delta(\vec r-\vec r_1)$ to verify that for $n\ge 2$
    \begin{eqnarray}
        Q'_{i_1 \cdots i_n} &=& -\frac{R}{r}\left.\vec p\cdot\nabla T_{i_1 \cdots i_n}(\vec u)\right|_{\vec u=\vec 0}\nn\\
        &=&0.
    \end{eqnarray}
 {For example, using \eqref{Tij} we verify that
    \begin{eqnarray*}\left.\nabla T_{ij}(\vec u)\right|_{\vec u=\vec 0}&=&\left[3u_i\nabla u_j+3u_j\nabla u_i-2u_k\nabla u_k\delta_{ij}\right]_{\vec u=\vec 0}\\
    &=&0.
    \end{eqnarray*}
    }
For higher-order multipoles, we establish our claim heuristically. Consider two equivalent but opposite multipoles of order \( n \), denoted as \( Q^{(1)}_{i_1,\dots,i_n} = Q_{i_1,\dots,i_n} \), located at \( \vec{r} \), and \( Q^{(2)}_{i_1,\dots,i_n} = -Q_{i_1,\dots,i_n} \), positioned at \( \vec{r} + \vec{a} \). In the limit \( |\vec{a}| \to 0 \), we assume that \( \|Q\|_1 a \) remains finite, where \( \|Q\|_1 \) represents the norm of the multipole, defined as
\[
\|Q\|_1 := \max{|Q_{i_1,\dots,i_n}|}.
\]
Under this limit, the two multipoles effectively combine to produce a multipole of order \( n+1 \), while higher-order contributions become negligible. The resulting image consists of a combination of multipoles of order \( m' \leq n \), with their loci separated by a distance
\[
\vec{a}\,' \approx \frac{R^2}{r^2} \vec{a}.
\]
For finite values of \( \vec{a} \), this combination retains multipoles of various orders; however, contributions of orders \( m' > n+1 \) vanish as \( |\vec{a}| \to 0 \).

\appendix
\section{Image of a dipole Parallel to the $ z $-axis}\label{App-parallel-dipole} 
We construct a dipole by placing a charge $q_1 = -q$ at $ \vec{r}_1 = d \hat{z} $ and another charge $ q_2 = +q $ at $ \vec{r}_2 = (d + a) \hat{z} $. The image charges $q'_1$ and $q'_2$ given in \eqref{parallel-q1} and \eqref{parallel-q2} are located, respectively, at
    \begin{align}
        &\vec s_1=\frac{R^2}{d}\hat z,&\vec s_2=\frac{R^2}{d+a}\hat z\simeq\frac{R^2}{d}\left(1-\frac{a}{d}\right)\hat z.
    \end{align}
Using \eqref{parallel-Q}, we verify that the total dipole moment with respect to the origin is 
    \bea
    \vec p\,'&=&\left(q'_1\vec s_1+q'_2\vec s_2\right)-Q\vec s_1\nn\\
    &=&\frac{R^3}{d^3}\vec p.
    \eea
This result can also be obtained by evaluating the dipole moment with respect to the image location:
    \bea
    \vec p\,'&=&q'_1\vec 0+q'_2(\vec s_2-\vec s_1)\nn\\
    &=&\frac{R^3}{d^3}\vec p.
    \eea

\section{Dipole-Induced Potential Near a Grounded Sphere}\label{App-dipole}
The electric potential of a dipole near a grounded sphere is given by \eqref{dipole-potential}. Here, we compute the contribution from  \eqref{dipole-potential-Q} and \eqref{dipole-potential-p}. 
    \bea
    \phi_Q(\vec r)+\phi_{p'}(\vec r) &:=& 
   \frac{1}{4\pi\epsilon_0}\left[ \frac{Q}{\left| \vec r - \frac{R^2}{d} \hat z \right|}
    + \frac{ \vec p\,' \cdot \left( \vec r - \frac{R^2}{d} \hat z \right) }
    { \left| \vec r - \frac{R^2}{d} \hat z \right|^3 }\right]\nn\\
    &=& \frac{1}{4\pi\epsilon_0}\left[\frac{p_2}{d\left|\frac{d}{R}\vec r - R \hat z\right| }
     + \frac{ p_2\left( \hat z\cdot\vec r-\frac{R^2}{d}\right)}{ \left|\frac{d}{R}\vec r - R \hat z\right|^3}-\frac{p_1  \hat x\cdot\vec r}{ \left|\frac{d}{R}\vec r - R \hat z\right|^3}\right] \nn\\
     \label{AppEq-dipole-1}
     &=&\frac{1}{4\pi\epsilon_0}\vec p\cdot\left(\frac{\frac{r^2}{R^2} \vec{r}_1 - \vec{r}}{\left| \frac{d}{R} \vec{r} - R \hat{z} \right|^3} \right),
    \eea
where we have used $\vec s_1=\frac{R^2}{d}\hat z$, $Q=\frac{Rp_2}{d^2}$,  $\vec p\,'=\frac{R^3}{d^3}\left(-p_1\hat x+p_2\hat z\right)$, $\vec p=p_1\hat x+p_2\hat z$, and $\vec r_1=d\,\hat z$.

To show that the potential \eqref{dipole-potential} satisfies the Dirichlet boundary condition, we use $\vec r_1=d\hat z$ and the identity 
    \bea
    \left| \frac{d}{R} \vec{R} - R \hat{z} \right|&=&\sqrt{d^2+R^2-2d\vec R\cdot\hat z}\nn\\
    &=&\left|\vec R-\vec r_1\right|.
    \eea
Plugging this result into \eqref{AppEq-dipole-1}, we verify that 
    \be 
    \phi_Q(\vec R)+\phi_{p'}(\vec R)=-\frac{1}{4\pi\epsilon_0}\vec p\cdot\left(\frac{\vec R-\vec r_1}{|\vec R-\vec r_1|}\right). 
    \ee
This confirms that the electric potential \eqref{dipole-potential} vanishes at $\vec r=\vec R$.
    
\section{The charge distribution of an ideal dipole}\label{Appendix-charge-distribution}
This appendix establishes the charge distribution of an ideal dipole, as given by~\eqref{dipole-density}. Recall that a dipole with moment $\vec{p}$ located at position $\vec{r}_1$ can be modeled as two point charges, $-q$ and $q$, placed at $\vec{r}_1$ and $\vec{r}_1 + \epsilon\,\hat{p}$, respectively,  where $\hat{p}$ denotes the unit vector along $\vec{p}$. As $\epsilon \to 0$ with $|\vec{p}| = q\epsilon$ held constant, the resulting configuration approaches the idealized dipole distribution. So, the corresponding charge distribution is given by 
    \bea
    \rho(\vec r)&=&\lim_{\epsilon\to0} q\left[-\delta(\vec r-\vec r_1)+\delta(\vec r-\vec r_1-\epsilon\,\hat p)\right]\nn\\
    &=&\lim_{\epsilon\to0} (-q\epsilon) \hat p\cdot\nabla\delta(\vec r-\vec r_1)\nn\\
    &=&-\vec p\cdot\nabla\delta(\vec r-\vec r_1). 
    \eea
To obtain the second equality, we have used the Taylor expansion of the Dirac delta function. 

This result can be confirmed as follows. Using this charge distribution in \eqref{monopole-moment}, we verify that the corresponding monopole moment is zero. Using it together with \eqref{Ti} in \eqref{multipole-moment}, the components of the dipole (in the notation of \eqref{multipole-moment}) can be obtained as follows.
    \bea
    Q_i&=&-\int d^3r \left[\vec p\cdot\nabla\delta(\vec y)\right]y_i\nn\\
    &=& \int d^3r \left[\vec p\cdot\nabla y_i\right]\delta(\vec y)\nn\\
    &=&p_i,
    \eea
where $p_i$ denotes the $i$-th component of the vector $\vec p$. Repeating this calculation with \eqref{Tij} or higher rank tensors $T_{i_1\cdots i_n}(\vec y)$, we observe that the corresponding quadrupole and higher order multipoles are zero. This confirms that \eqref{dipole-density} gives the charge distribution of an ideal dipole. 

\section{Green's Function Analysis of Dipole-Induced Potential Near a Grounded Sphere}\label{App-Green-dipole}
To calculate the electric potential \eqref{dipole-potential-Green}, we need the following equations. 
\bea
\nabla' G(\vec r,\vec r\,')  
&=& \nabla' \left( \frac{1}{|\vec r - \vec r\,'|} - \frac{R}{r' \left|\vec r - \frac{R^2}{r'^2}\vec r\,'\right|} \right) \nn\\
&=&  \frac{\vec r-\vec r\,'}{|\vec r-\vec r\ '|^3}  - \nabla' \left( \frac{R}{\sqrt{r^2 r'^2 + R^4 - 2 R^2\vec r\cdot\vec r\,'}} \right). 
\eea 
where $r:=|\vec r|$ and $r':=|\vec r\,'|$. Furthermore,
\bea 
\nabla' \left( \frac{R}{\sqrt{r^2 r'^2 + R^4 - 2 R^2\vec r\cdot\vec r\,'}}\right)&=& - R  \frac{r^2\vec r\,'-R^2\vec r}{\left( r^2 r'^2 + R^4 - 2 R^2\vec r\cdot\vec r\,'\right)^{3/2}} \nn\\
&=& - \frac{\frac{r^2}{R^2}\vec r\,'-\vec r}{\left( \frac{r^2 r'^2}{R^2} + R^2 - 2 \vec r\cdot\vec r\,'\right)^{3/2}}\nn\\&=&
-\frac{\frac{r^2}{R^2}\vec r\,'-\vec r}{|\frac{r'}{R}\vec r-R\hat{r'}|^3},
\eea 
where $\hat {r'}:=\vec r\,'/r'$. Using these results in  \eqref{dipole-potential-Green}, we get \eqref{dipole-potential}. 

\section{Images of a quadrupole}\label{App-images}
Using \eqref{dq}, \eqref{dq'}, \eqref{image-charge-general}, and \eqref{quadrupole-density} we obtain
\bea
q' &=& - R \int d^3 r \, \frac{\rho(\vec{r})}{r}\nn\\
&=& - \frac{1}{6} R Q_{ij} \int d^3 r \, \delta(\vec{r} - \vec{r}_1) \, \partial_i \partial_j \left( \frac{1}{r} \right)\nn\\
&=& - \frac{1}{2} R Q_{ij} \frac{(r_1)_i (r_1)_j}{|\vec r_1|^5},
\eea
where the second equality is obtained by integration by parts. To obtain the third equality we have noted that $\sum_{i=1}^3Q_{ii}=0$. Similarly, \eqref{image-dipole-general} gives
    \bea
    \vec{p}\,' 
    &=& -\frac{1}{6} R^3 Q_{ij} \int d^3 r \, \delta(\vec r - \vec r_1) \, \partial_i \partial_j \left[ \frac{1}{r} \left( \frac{\vec r}{r^2} - \frac{\vec r_1}{|\vec r_1|^2} \right) \right]\nn\\
    &=& -\frac{1}{2} R^3 Q_{ij} \frac{4 (r_1)_i (r_1)_j \vec r_1- |\vec r_1|^2 \left( (r_1)_i \hat{e}_j + (r_1)_j\hat{e}_i \right)}{|\vec r_1|^7}\nn\\
    &=&R^3 Q_{ij} \frac{(r_1)_i}{|\vec r_1|^5} \hat{e}_j + \frac{4 R^2}{|\vec{r}_1|^2} q' \vec r_1.
    \eea
Using \eqref{image-quad-general} we obtain 
    \bea 
    Q'_{ij} &=& -\frac{1}{2} R^5{Q_{kl}} \left\{ \partial_i \partial_j \left[\frac{1}{r} \left( \frac{x_k}{r^2} - \frac{(r_1)_k}{|\vec r_1|^2} \right) \left( \frac{x_l}{r^2} - \frac{(r_1)_l}{|\vec r_1|^2} \right) \right] \right\}_{\vec r = \vec r_1}
    \nn\\&=&-R^5{Q_{kl}} \left\{ \frac{1}{r} \left[ \partial_i \left( \frac{x_k}{r^2} - \frac{(r_1)_k}{|\vec r_1|^2} \right) \partial_j \left( \frac{x_l}{r^2} - \frac{(r_1)_l}{|\vec r_1|^2} \right)\right] \right\}_{\vec r = \vec r_1}
      \nn\\&=& -R^5{Q_{kl}}  \frac{1}{|\vec r_1|} \left( \frac{\delta_{ik}}{|\vec r_1|^2} - \frac{2 (r_1)_i (r_1)_k}{|\vec r_1|^4} \right) \left( \frac{\delta_{jl}}{|\vec r_1|^2} - \frac{2 (r_1)_j (r_1)_l}{|\vec r_1|^4} \right) \nn\\&=& 
      -\frac{R^5}{|\vec r_1|^5}{Q_{ij}} + \frac{2 R^5}{|\vec r_1|^7} \Big( Q_{ik} (r_1)_k (r_1)_j + i \leftrightarrow j \Big) - \frac{4 R^5}{|\vec r_1|^9} \Big( Q_{kl} (r_1)_k (r_1)_l \Big)(r_1)_i (r_1)_j.
    \eea 
\section{Derivation of the Quadrupole Tensor \eqref{example 1}}\label{App-Quad-Example1}
The quadrupole tensor is given by \cite{Jackson}
\be 
Q_{ij} = \int d^3r\left(3x_i x_j - r^2 \delta_{ij}\right) \rho(\vec{r}),
\ee
where $\rho(\vec{r})$ is the charge distribution, and $x_i$ is the $i$-the component of the $\vec r$. In Example 1, we consider
    \be 
    \rho(\vec r)=q\,\delta(z-d)\left\{\Big[\delta(x-a)+\delta(x+a)\Big]\delta(y)-\delta(x)\Big[\delta(y-a)+\delta(y+a)\Big]\right\}.
    \ee
Since $\rho(x_1,x_2,x_3)=\rho(-x_1,x_2,x_3)$, it follows that  
    \be 
    \int d^3r\,x_1 x_j \rho(\vec{r})=0,\qquad \mbox{for}\qquad j=2,3.
    \ee
This shows that $Q_{1j}=0$ for $j=2,3$. Similarly, symmetry under $x_2\to-x_2$ yields $Q_{2j}=0$ for $j=1,3$. Consequently
    \be 
    Q_{ij}=0 \qquad \mbox{for}\qquad i\neq j.
    \ee
The diagonal components can be computed directly. 
    \bea 
    Q_{11}&=&\int d^3r\left(2x_1^2-x_2^2-x_3^2\right) \rho(\vec{r})\nn\\
    &=&q\left\{\Big[(2a^2-d^2)+(2a^2-d^2)\Big]-\Big[(-a^2-d^2)+(-a^2-d^2)\Big]\right\}\nn\\
    &=&6qa^2,\\
    Q_{22}&=&\int d^3r\left(-x_1^2+2x_2^2-x_3^2\right) \rho(\vec{r})\nn\\
    &=&q\left\{\Big[(-a^2-d^2)+(-a^2-d^2)\Big]-\Big[(2a^2-d^2)+(2a^2-d^2)\Big]\right\}\nn\\
    &=&-6qa^2,\\
    Q_{33}&=&\int d^3r\left(-x_1^2-x_2^2+2x_3^2\right) \rho(\vec{r})\nn\\
    &=&q\left\{\Big[(-a^2-d^2)+(-a^2-d^2)\Big]-\Big[(-a^2-d^2)+(-a^2-d^2)\Big]\right\}\nn\\
    &=&0.
    \eea 

\section{Green's Function Analysis of the electric potential of the image quadrupole}\label{App-Green-image-quadrupole}
We use the Green's function \eqref{Green-image} and the charge distribution \eqref{quadrupole-density} to obtain \eqref{image-quadrupole-potential}. We have 
    \be 
    \label{App-1}
    \phi_{\rm im}(\vec{r}) = \int d^3 r' \, \rho(\vec{r}\,')\, G_{\rm im}(\vec{r}, \vec{r}\,')  
= \frac{1}{6} Q_{ij} \left[ \frac{\partial^2}{\partial y_i \partial y_j} G_{\rm im}(\vec{r}, \vec{y}) \right]_{\vec{y} = \vec{r}_1},
    \ee 
where $y_i:=\vec y\cdot\hat e_i$
To compute this we note that 
    \be 
    \frac{\partial}{\partial y_i} \left( \frac{R/y}{|\vec{r} - \frac{R^2}{y^2} \vec{y}|} \right)
= \frac{R(R^2x_i - r^2y_i)}{\left( y^2 r^2 + R^4 - 2 R^2 \vec{y} \cdot \vec{r} \right)^{3/2}},
    \ee 
where $y:=|\vec y|$. Using this result, we obtain
    \be 
    \frac{\partial^2}{\partial y_i \partial y_j} \left( \frac{R /y}{|\vec{x} -\frac{R^2}{y^2} \vec{y}|} \right)
= \frac{3R(R^2x_i - r^2 y_i)(R^2 x_j - r^2 y_j)}{\left( y^2 r^2 + R^4 - 2R^2 \vec{y} \cdot \vec{r} \right)^{5/2}}
- \frac{R r^2 \delta_{ij}}{\left( y^2 r^2 + R^4 - 2R^2 \vec{y} \cdot \vec{r} \right)^{3/2}}.
    \ee 
Using this result in \eqref{App-1}, and noting that $Q_{ij}\delta_{ij}=0$, we confirm \eqref{image-quadrupole-potential}.

\end{document}